\documentclass[twocolumn,showpacs,preprintnumbers,amsmath,amssymb]{revtex4}

\usepackage{epsfig,psfrag}
\usepackage{dcolumn}
\usepackage{bm}
\usepackage{graphicx}
\usepackage{color}

\begin{document}

\title{Atom-dimer scattering for confined ultracold fermion gases}

\author{C.~Mora,$^1$ R.~Egger,$^1$  A.O.~Gogolin,$^2$ and A.~Komnik$^3$}

\affiliation{${}^1$~Institut f\"ur Theoretische Physik, 
Heinrich-Heine-Universit\"at,
 D-40225 D\"usseldorf, Germany\\
${}^{2}$ Department of Mathematics, Imperial College, 180 Queen's Gate,
London SW7 2BZ, United Kingdom \\
${}^{3}$  Physikalisches Institut, Albert-Ludwigs-Universit\"at, D-79104 
Freiburg, Germany
}

\date{\today}

\begin{abstract}
We solve the three-body problem of a quasi-one-dimensional
ultracold Fermi gas with parabolic confinement length $a_\perp$ and 
3D scattering length $a$. On the two-body level, there is a Feshbach-type
resonance at $a_\perp/a\approx 1.46$, and a dimer state for 
arbitrary $a_\perp/a$.  The three-body problem is shown to be universal, and
described by the atom-dimer scattering length $a_{ad}$ 
and a range parameter $b_{ad}$.
In the dimer limit $a_\perp/a\gg 1$,  we find
a repulsive zero-range atom-dimer interaction.
For $a_\perp/a\ll -1$, however, the potential has long range,
with $a_{ad}>0$ and $b_{ad}\gg a_{ad}$. There is no trimer state, and
despite $a_{ad}=0$ at $a_\perp/a\approx 2.6$, there is 
no resonance enhancement of the interaction.
\end{abstract}


\maketitle

The recent experimental observation of the formation of dimers (molecules) 
in ultracold binary Fermi gases \cite{molecule1} has sparked intense 
excitement and activity among both atomic and condensed-matter physicists.
Experiments are now able to probe in detail the BEC-BCS crossover regime
by using magnetic-field tuned Feshbach resonances
 \cite{regal04,grimm04,ketterle04},
where the 3D scattering length $a$ describing the $s$-wave
interaction strength among different fermion species can be tuned
almost at will.  This crossover remains a difficult 
and long-standing challenge to theory, as there
is no small parameter in the  problem \cite{timmermanns}.
Here we discuss the related but simpler problem of an ultracold
two-species ($\uparrow,\downarrow$)
 Fermi gas confined in a 1D trap potential. On the two-body level,
there is always a bound state (dimer), and one has a confinement-induced
resonance (CIR) in the 1D atom-atom scattering length $a_{aa}$
\cite{olshanii98,bergeman03},
similar to a Feshbach resonance. 
Such a scenario appears to be experimentally feasible 
\cite{molecule1,regal04,grimm04,ketterle04,gorlitz,paredes}
and could reveal interesting new physics.
In this paper, we analytically solve the three-body problem  
for confined ultracold fermions,
and compute the atom-dimer scattering length $a_{ad}$ 
and the potential range $b_{ad}$ [see Eq.~(\ref{aaddef}) below].
Our results also determine parameters entering models
that may be solved exactly by powerful many-body techniques
in 1D \cite{book}, see also Refs.~\cite{tokatly,fuchs}.

For simplicity, we assume the two fermion species to have the same mass $m_0$.
Under the harmonic transverse confinement
potential $U_c({\bf r}) = \frac12 m_0\omega_\perp^2 (
x^2+y^2)$,  with associated lengthscale 
$a_\perp=(2\hbar/m_0\omega_\perp)^{1/2}$, there is
a two-body bound state (dimer) 
with dimensionless binding energy 
\begin{equation}\label{omegab}
\Omega_B = (\hbar\omega_\perp-E_B)/2\hbar\omega_\perp
\end{equation}
 determined by the condition
\cite{olshanii98,bergeman03}
\begin{equation}\label{bs}
\zeta(1/2,\Omega_B) + a_\perp/a =0.
\end{equation}
Since the zeta function $\zeta(1/2,\Omega)$ is monotonic in $\Omega$, there is 
exactly one bound state for any given $a_\perp/a$, although the
3D problem has a bound state only for $a>0$. 
For $a_\perp/a\to -\infty$, the `BCS limit' is reached, where
$\Omega_B\simeq (a/a_\perp)^2 \ll 1$, and the dimer size is large,
of the order $a_\perp^2/|a|$.  
In the tightly bound `dimer limit', $a_\perp/a\to +\infty$, 
the dimer size is small, of the order
$a$, and $\Omega_B\simeq (a_\perp/2a)^2\gg 1$.
The analogue of the Feshbach resonance is then realized by the 
CIR.  Solving the two-body scattering problem with just one open channel,
the 1D scattering length is  \cite{olshanii98,bergeman03}
\begin{equation}\label{a1d}
a_{aa}=-\frac{a_\perp}{2} \left[\frac{a_\perp}{a}-{\cal C}\right],\quad
{\cal C}=-\zeta(1/2)\simeq 1.4603.
\end{equation}
At low energies, this implies that one can use
the  1D atom-atom interaction potential 
$V_{aa}(z,z')=g_{aa} \delta(z-z')$ with $g_{aa}=-2\hbar^2/m_0 a_{aa}$.
The CIR then occurs for $a_{aa}=0$, corresponding to $\Omega_B=1$,
and can be reached by tuning $a_\perp$ or $a$.

A natural question then concerns the scattering properties of
the atom-dimer system, for instance, the scattering length
$a_{ad}$.  This problem has recently been solved for 
the unconfined 3D case by Petrov \cite{petrov03},
where $a_{ad}\approx 1.2 a$.  In the confined
geometry, it is then interesting to ask (i) whether the 
scattering length $a_{ad}$ also shows CIR-related resonant behavior, (ii)
whether a trimer state could be possible in the confined geometry, and 
(iii) whether a universal description in terms of 
simple two-body physics is always applicable.
For the unconfined case, Efimov \cite{efimov} showed that this
three-body problem is universal.  Moreover, there is no bound trimer state. 
Below we shall answer these questions in detail.

We study the three-body problem $(\uparrow\uparrow\downarrow)$ with
two identical fermions, and denote by ${\bf x_1}$ $({\bf x_{2,3}})$ the 
 position of the $\downarrow$ (the two $\uparrow$) particles.
Next we perform an
orthogonal transformation to variables $({\bf x}, {\bf y}
,{\bf z})$ in order to decouple the center-of-mass coordinate ${\bf z}$
\cite{petrov03}.  Since the confinement is harmonic, $U_c$ remains 
diagonal in the positions.
The three-body problem then reduces to  
\begin{equation}\label{schro}
 \left( -\frac{\hbar^2}{m_0} \nabla_{\bf X}^2 + U_c({\bf X}) -E \right)
 \Psi({\bf X}) = - \sum_{\pm}  V ( {\bf r}_{\pm} )  \Psi({\bf X}),
\end{equation}
where ${\bf X} = ({\bf x},{\bf y})$ is a six-dimensional vector,
${\bf x}=(2{\bf x}_1-{\bf x}_2-{\bf x}_3)/2$ and 
${\bf y}={\bf x}_3-{\bf x}_2$.
With these definitions, the distances between 
the $\downarrow$ particle and each $\uparrow$ particle
are ${\bf r}_{\pm} = \sqrt{3}{\bf x}/2 \pm  {\bf y}/2$.
We adopt the pseudopotential approximation for the
 3D interaction,
 $V({\bf r})=(4\pi\hbar^2 a/m_0)\delta({\bf r}) 
\frac{\partial}{\partial r} (r \cdot)$, which 
allows to incorporate interactions  via
boundary conditions imposed for vanishing distances between $\uparrow$ and
$\downarrow$ atoms.  For ${\bf r}_{\pm} \to 0$, this implies the
singular behavior
\begin{equation}\label{asymto}
\Psi({\bf X}) \simeq \mp \frac{f({\bf r}_{\perp,\pm})}{4 \pi r_\pm} (1-
r_\pm/a ),
\end{equation}
where the ${\bf r}_{\perp,\pm} =  {\bf x} /2 \mp 
\sqrt{3} {\bf y}/2$ are orthogonal to ${\bf r}_{\pm}$.
We consider $E<0$ and write 
\[
E = -2\Omega_B \hbar \omega_\perp + \hbar^2\bar{k}^2/m_0,
\]
where the relative momentum $ \hbar\bar{k}$ of the atom-dimer complex
is sent to zero later.

The boundary conditions (\ref{asymto})
allow to write Eq.~(\ref{schro}) in the form
\begin{equation}\label{gen2}
m_0\Psi({\bf X})  = \sum_{\pm} \mp
 \int d {\bf x'} d {\bf y'}
G_E^{(2)} ({\bf x},{\bf y};{\bf x}',{\bf y}') 
f({\bf r}'_{\perp,\pm}) 
\delta( {\bf r}'_{\pm} )  ,
\end{equation}
where the two-particle Green function is
\[
G_E^{(2)} ( {\bf r}_1,{\bf r}_2, {\bf r}_3, {\bf r}_4) 
= \sum_{\lambda_1,\lambda_2} \frac{\psi_{\lambda_1} (  {\bf r}_1 )
\psi_{\lambda_2} (  {\bf r}_2 ) \psi_{\lambda_1}^* (  {\bf r}_3 )
\psi_{\lambda_2}^* (  {\bf r}_4 )}{E_{\lambda_1} + E_{\lambda_2} -  E}.
\]
Here $\psi_\lambda$ denotes the eigenfunctions to the single-particle
problem for eigenenergy $E_\lambda$. The quantum
numbers $\lambda$ include the 1D momentum $k$, the (integer)
angular momentum $m$,
and the radial quantum number $n=0,1,2\ldots$, which gives
$E_\lambda=\hbar\omega_\perp(2n+|m|+1) +\hbar^2 k^2/m_0$ and
$\psi_\lambda= e^{im\phi}R_{nm}(\rho)e^{ikz}$ 
with radial functions $R_{nm}$ \cite{olshanii98}.
Using the fact that $G_E^{(2)}$ and the integration measure are
both invariant under orthogonal transformations,
we find from Eq.~(\ref{gen2})
\begin{eqnarray*}
&& m_0 \Psi ( {\bf r} ,  {\bf r}_\perp) =  \int d {\bf r}^\prime_\perp  \,
f({\bf r}_\perp^\prime)  [
G_E^{(2)}( {\bf r}  , {\bf r}_\perp; 0,{\bf r}_\perp^\prime)  \\
&& - G_E^{(2)} ({\bf r}/2 + \sqrt{3} {\bf r}_\perp /2 , \sqrt{3}  {\bf r}/2
-{\bf r}_\perp /2; 0 ,{\bf r}^\prime_\perp) ],
\end{eqnarray*}
where $ {\bf r} \equiv  {\bf r}_{-}$ and ${\bf r}_\perp
\equiv {\bf r}_{\perp,-}$.
Next we implement the ${\bf r} \to 0$ limit according to Eq.~(\ref{asymto})
to obtain a closed equation for $f({\bf r}_{\perp})$.
This limit can be directly taken for the non-singular second term 
in the integral above,
while the first term contains the  singular behavior necessary from
 Eq.~(\ref{asymto}).
Once this singular behavior is removed, one obtains a regular
integral equation for $f({\bf r}_{\perp})$ 
\cite{petrov03,skorniakov,petrov04}.  It is convenient to
transform from real space into the complete basis $\{ \psi_\lambda \}$, 
implying 
\begin{equation}\label{integral-equa}
{\cal L}(\Omega_\lambda) f_\lambda = 
\sum_{\lambda'} A_{\lambda,\lambda'} \, f_{\lambda'},
\end{equation}
where we use [see also Eq.~(\ref{bs})]
\begin{eqnarray*}
\mathcal{L}(\Omega) & =&  \zeta(1/2, \Omega) - \zeta(1/2,\Omega_B) ,\\
\Omega_\lambda & = & \Omega_B - (a_\perp \bar k/2)^2
+ E_\lambda/2 \hbar \omega_\perp,
\end{eqnarray*}
and the matrix
$A_{\lambda,\lambda'}$ is given by
\[
\frac{4\pi a_\perp}{m_0}
\int d {\bf r}_\perp \, \psi_{\lambda}^* ( {\bf r}_\perp)\,
\psi_{\lambda'} (- {\bf r}_\perp /2) 
G_{E-E_\lambda} (  \sqrt{3} {\bf r}_\perp/2,0 ).
\]
Using the integral representation of Ref.~\cite{tokatly}
for the two-body Green function
$G_E({\bf r},{\bf r}')$,  $A_{\lambda,\lambda'}$ can be evaluated
explicitly. 
Before analyzing Eq.~(\ref{integral-equa}) further, however, it is useful
to perform a rescaling. 
So far, $f$ has been taken as a function of ${\bf r}_\perp=(\rho,z)$.
However, for the asymptotic solution consisting of
a dimer and a free atom, the atom-dimer distance ${\bf d}$ does not 
coincide with ${\bf r}_\perp$.  With the dimer wavefunction
$\Phi_0({\bf r})$ \cite{olshanii98},
we expect 
\[
\Psi({\bf r},{\bf r}_\perp)\simeq \Phi_0({\bf r}) \chi({\bf d}),\quad
{\bf d}=({\bf x}_1+{\bf x}_3)/2-{\bf x}_2={\bf r}_+-{\bf r}_-.
\]  
In the $r\to 0$ limit, $\Psi \simeq \chi(\sqrt{3} {\bf r}_\perp/2 ) 
/4\pi r$, which establishes
a connection between ${\bf d}$ and ${\bf r}_\perp$ from
Eq.~(\ref{asymto}).
After the rescaling ${\bf r}_\perp\to \sqrt{3} {\bf r}_\perp/2$,
$f$ coincides with the scattering solution $\chi$. 
Therefore, from now on, all
wavevectors are rescaled by the factor $2/\sqrt{3}$. 

Let us then proceed by projecting the integral equation
 (\ref{integral-equa}) to the lowest transverse state ($n=m=0$).
Explicit calculations \cite{future} 
show that the higher states are negligible in
the BCS limit and affect $a_{ad}$ only slightly in the dimer limit,
see below.
Taking into account the above rescaling, noting that only $m=0$
modes have nonzero overlap with the lowest state
we find
\begin{eqnarray}\label{equa}
{\cal L}(\Omega_k) f_k 
&=& \int_{-\infty}^{+\infty} \frac{d k'}{2 \pi}  A_{k,k'}  f_{k'},\\
\nonumber
\Omega_k &=&\Omega_B +\frac{ 3a_\perp^2(k^2-\bar{k}^2)}{16}
\end{eqnarray}
with the kernel given by
\begin{equation}\label{deltakk}
A_{k,k' }  =  \sum_{p=0}^{\infty}  
 \frac{4^{-p}}{p+ \Omega_B + (a_\perp/2)^2[- 3 \bar{k}^2 /4
+ k^2 + k'^2 + k k'] } .
\end{equation}

Following Ref.~\cite{skorniakov}, we now make an {\sl ansatz} for the
solution of the integral equation, 
\begin{equation}\label{skor}
f (k) = 2 \pi \delta( k-\bar{k} ) + i \tilde{f} (k,\bar{k} ) \sum_\pm
\frac{1}{\bar{k} \pm k + i 0^+}, 
\end{equation}
with a regular function $\tilde{f}(k,\bar{k})$. This
ansatz gives the expected scattering state after Fourier 
transforming to real space, 
\[
f(z) = e^{i\bar{k} z}  +  \tilde{ f}
\left ( {\rm sgn}(z) \bar{k}, \bar{k}\right ) e^{i\bar{k} |z|}.
\]
In the low-energy limit $k,\bar{k}\to 0$, 
on general grounds \cite{landau}, the expansion 
\begin{equation} \label{aaddef}
\tilde{f}
(k,\bar{k}) = -1 + i k b_{ad} + i\bar{k} a_{ad} + {\cal O}(k^2, \bar{k}^2,
k \bar k)
\end{equation}
applies, where $a_{ad}$ is the atom-dimer 
scattering length.  From the analysis
of model potentials, the length $b_{ad}$ is 
linked to the range of the effective
1D atom-dimer potential. In particular, $b_{ad}\to 0 $ if a
zero-range $\delta$-potential can be used for the effective 1D 
atom-dimer scattering at low energy scales. 

Inserting the ansatz (\ref{skor}) into Eq.~(\ref{equa}), we obtain
\begin{eqnarray}
\nonumber
&& \frac{\mathcal{L} ( \Omega_{k} )}{\bar{k}^2 - k^2} 2 i \bar{k}
\tilde{f}(k,\bar{k}) - i \mathcal{P} \sum_\pm
\int_{-\infty}^{+\infty} \frac{d k'}{2 \pi}
 \frac{A_{k,k'}}{\bar{k} \pm k'}  
 \tilde{f} (k',\bar{k}) \\  
\label{eqint2}
&& - \frac12 \left[ \tilde{f} (\bar{k},\bar{k})   A_{k,\bar{k}}+ 
\tilde{f} (-\bar{k},\bar{k})   A_{k,-\bar{k}} \right] = A_{k,\bar{k}},
\end{eqnarray}
where ${\cal P}$ denotes a principal value integration.
Finally, the analysis can be simplified considerably
by letting $\bar{k} \to 0$, i.e., by expanding Eq.~(\ref{eqint2})
in $\bar{k}$ and keeping only the lowest order.
At that stage, we switch to dimensionless momenta $u,u'$ by writing
$k=(2\sqrt{\Omega_B}/a_\perp) u$. Some algebra gives,
with the weakly $\Omega_B$-dependent functions
\begin{eqnarray*}
G(u,u')&=& \sum_{p=0}^\infty \frac{4^{-p}}{1+u^2+u^{\prime 2}+ 
u u^\prime + p/\Omega_B},\\
H(u) &=& \sum_{p=0}^\infty \frac{4^{-p} u}{2(1+u^2+p/\Omega_B)^{-2} },
\end{eqnarray*}
the following integral equation 
for $h(u)\equiv \tilde{f}(u,0)$:
\begin{eqnarray} \nonumber
&& \int_{-\infty}^{+\infty} \frac{d u'}{ 2\pi u'^2} 
[ G(u,u') h(u') -  G(u,0) h(0)]  
\\ \label{inteqfin}
&& - \frac{ \sqrt{\Omega_B}}{2u^2} 
{\cal L}\left(\Omega_B\left[1+\frac{3u^2}{4}\right]\right) h(u)
\\ \nonumber &&= \frac{a_{ad}\sqrt{\Omega_B}}{a_\perp} G(u,0) +iH(u). 
\end{eqnarray}
Note that the real (imaginary) part of $h(u)$ is even (odd) in $u$.
The scattering length $a_{ad}$ finally follows from 
the real part of Eq.~(\ref{inteqfin}) and the condition $h(0)=-1$,
see Eq.~(\ref{aaddef}), while $b_{ad}$ can be extracted from the
imaginary part of Eq.~(\ref{inteqfin}).
The integral equation (\ref{inteqfin})
 shows in particular that $a_{ad}/a_\perp$ depends only on $\Omega_B$, 
and hence only on the binding energy of the dimer.

In the {\sl dimer limit}, $a_\perp/a\gg 1$ and $\Omega_B\gg 1$,  see
Eq.~(\ref{bs}), power counting 
shows that the first term in Eq.~(\ref{inteqfin}) is negligible.
Using $\zeta(1/2,\Omega\gg 1)\approx -2\sqrt{\Omega}$ and $G(0,0)=4/3$
yields from $h(0)=-1$ the result 
\begin{eqnarray} \label{dimer1}
a_{ad} &=& - \kappa_\infty \sqrt{\Omega_B} a_\perp+ \beta a_\perp
 /\sqrt{\Omega_B} \\
\nonumber &=&  - \kappa_\infty a_\perp^2/2a  + 2 \beta a,
\end{eqnarray}
where $\kappa_\infty=9/32=0.28125$.  We also specify the subleading
order in Eq.~(\ref{dimer1}), where $\beta \approx 0.543$. 
A similar calculation gives $b_{ad}/a_\perp = (8/9)\Omega^{-3/2}_B$ 
for $\Omega_B\gg 1$, validating a repulsive zero-range 1D
atom-dimer potential in the low-energy limit,
$V_{ad}(z)=g_{ad} \delta(z)$ with $g_{ad}>0$.
Due to the presence of higher channels, $a_{ad}$ is renormalized
in the dimer limit. This correction can be derived analytically \cite{future}
by making explicit contact to the integral equation in the unconfined
case \cite{petrov03,skorniakov}. We find $\kappa_\infty\approx 0.636$, which
gives $a_{ad} = -(a^r_\perp)^2/2 \, (1.2\,a)$ with 
the confinement scale $a_\perp^r=
(3 \hbar/2 m_0 \omega_\perp)^{1/2}$ for the atom-dimer reduced mass 
$2 m_0/3$.  With the 3D atom-dimer scattering length $1.2\, a$ \cite{petrov03},
this result exactly matches the 
dimer limit of the analogous two-body result (\ref{a1d}).
Since closed channels do not cause
profound changes even in the dimer limit, 
they are neglected in what follows.

\begin{figure}
\scalebox{0.65}{
\includegraphics{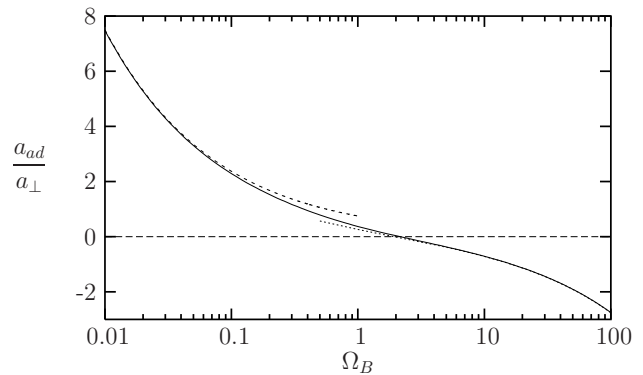}
}
\caption{\label{fig1} Scattering length $a_{ad}/a_\perp$ versus dimensionless
binding energy $\Omega_B$. The solid curve is the numerical solution to
Eq.~(\ref{inteqfin}), and the dotted (dashed) curves represent the 
analytical results in the dimer (BCS) limit, respectively,
see Eqs.~(\ref{dimer1}) and  (\ref{bcs1}).}
\end{figure}
\begin{figure}
\scalebox{0.65}{
\includegraphics{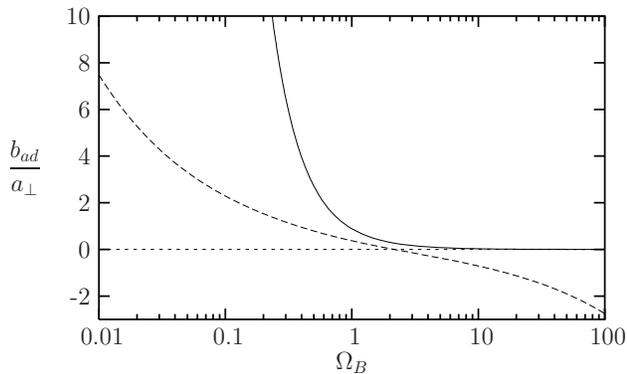}
}
\caption{\label{fig2} Range parameter $b_{ad}/a_\perp$ versus
$\Omega_B$. The solid curve gives $b_{ad}$ from the numerical solution of
Eq.~(\ref{inteqfin}), and the dashed line represents $a_{ad}$ from
Fig.~\ref{fig1} for comparison.
}
\end{figure}

Outside the dimer limit, in general a numerical solution
of Eq.~(\ref{inteqfin}) is necessary. To ensure regularity
of $h(u)$, it is beneficial to Fourier transform back to real space,
where the Fourier transformed $h(z)$ is well behaved and allows for 
a quickly converging solution of the integral equation. 
The numerical result for $a_{ad}/a_\perp$ as function
of $\Omega_B$ is shown in Figure \ref{fig1}.
In the {\sl BCS limit}, where $a_\perp/a\ll -1$ and $\Omega_B\ll 1$, 
we find 
\begin{equation}\label{bcs1}
a_{ad} = \kappa_0 a_\perp /\sqrt{\Omega_B} = \kappa_0 a_\perp^2/|a| ,
\quad \kappa_0\approx 0.75.
\end{equation}
The solution in Fig.~\ref{fig1} for arbitrary $\Omega_B$
nicely matches onto the limits (\ref{dimer1}) and (\ref{bcs1}).
Remarkably, around $\Omega_B\approx 2.2$, there is a zero of $a_{ad}$.
At first sight, this behavior seems to be linked to the
two-body CIR, see Eq.~(\ref{a1d}). However, the atom-dimer `resonance'
occurs at a different $\Omega_B$, and, more importantly, the
assumption of a $\delta$-potential interaction breaks down.
This can be seen by computing the range parameter $b_{ad}$ in
Eq.~(\ref{aaddef}), see Figure \ref{fig2}.  
While in the dimer limit, $b_{ad}$ stays small,
in accordance with our analytical result, in the 
BCS limit, it is found to diverge as $b_{ad}\propto \Omega_B^{-3/2}$. 
This implies that one 
cannot use a $\delta$-potential to describe atom-dimer scattering
outside the dimer limit, but instead more complicated potentials
have to be used, e.g.,  a repulsive square-well potential.
Furthermore, the potential becomes non-local \cite{footnote}.
An effective 1D potential for low-energy atom-dimer scattering can be 
constructed directly from Eq.~(\ref{equa}).
In fact, transforming Eq.~(\ref{equa}) back to real space, one gets a
Schr\"odinger equation
\begin{equation}
\left(-\frac{d^2}{dz^2} -\bar{k}^2 \right) f(z) = - \int dz'
V_{ad}(z,z') f(z')
\end{equation}
with the {\sl non-local} effective potential 
\begin{equation}
V_{ad}(z,z') = -\int \frac{dk}{2\pi} \frac{dk'}{2\pi} e^{ikz-ik'z'}
\frac{k^2-\bar{k}^2} {{\cal L}(\Omega_k)} A_{k,k'} ,
\end{equation}
where $A_{k,k'}$ is given by Eq.~(\ref{deltakk}).
It can be checked easily that
this potential becomes very wide in the BCS limit, with support
given by the dimer size, but it always stays 
repulsive.   Hence there is {\sl no
three-body bound state} (trimer) 
even in the confined problem for arbitrary $a_\perp/a$.

Finally, we have also analyzed the problem of three-body
recombination and dissocation, $\uparrow+\uparrow+\downarrow\leftrightarrow
 \uparrow \downarrow +\uparrow$, see Ref.~\cite{petrov03} for the
unconfined case, where a non-zero incoming state $\Psi_0$  
has to be taken into account in Eq.~(\ref{gen2}). 
 Antisymmetry of $\Psi_0({\bf x},
{\bf y})$ under ${\bf y}\to -{\bf y}$ requires that the 
lowest transverse state (for ${\bf y}$) has angular momentum $m=\pm 1$,
implying that dissociation processes encounter 
an energy barrier $\hbar\omega_\perp
(1+\Omega_B)$ not present in the unconfined problem.  
The three-body recombination problem is therefore simpler, and
dimers are rather stable against three-body
dissociation.

To conclude, we have solved the three-body problem of  a binary
cold Fermi gas.  The scattering length $a_{ad}$ describing atom-dimer
scattering has been extracted for arbitrary confinement and
3D scattering length, and is a universal function of $a_\perp/a$.
For $\Omega_B\approx 2.2$, corresponding to $a_\perp/a\approx 2.6$,
we find $a_{ad}=0$.  However, this does not imply resonantly
enhanced atom-dimer interactions.
While in the dimer limit,
$a_\perp/a\gg 1$, a standard repulsive zero-range potential is found,
in the BCS limit, $a_\perp/a\ll -1$, the situation turns out to be
more complicated.  The scattering
length is positive, $a_{ad}>0$, 
but at the same time the range of the effective
interaction becomes very large,  $b_{ad}\gg a_{ad}$.
Therefore it is not possible to employ zero-range potentials in that
limit anymore.  
Nevertheless, it is worth pointing out that
also the  confined three-body problem is
{\sl universal}\ in the sense that it can be completely expressed in
terms of two-body quantities.   This is encouraging news
for many-body calculations which rely on model parameters
extracted only from two-body physics \cite{tokatly,fuchs}.
Finally, we mention that the scattering length $a_{ad}$ (and
possibly the length $b_{ad}$) can be extracted experimentally
using standard techniques, and thus the scenario put forward
above could be checked in detail.
Future work should also address the role of Pauli blocking due
to the background Fermi sea. For the unconfined case,
this was studied by Combescot \cite{combescot03}.  
Furthermore, in a similar fashion as done here, dimer-dimer
scattering in the confined geometry can be analyzed from the four-body problem.
While in the deep dimer limit, contact to the free-space 
result \cite{petrov04} 
can again be established, the behavior for $\Omega_B\alt 1$ is 
particularly noteworthy.  This will be discussed elsewhere
\cite{future}. 

This work was supported by the SFB TR12 of the DFG.


\begin{references}
\bibitem{footnote} 
For a local potential, the sign of $b_{ad}$ is negative, while
for the non-local potential encountered here, the exact solution
gives a positive sign.
\end{references}
\end{document}